\documentclass[aps, prb, twocolumn, superscriptaddress]{revtex4}

\usepackage{amstext}
\usepackage{amsmath}
\usepackage{amssymb}
\usepackage{amsfonts}
\usepackage{graphicx}
\usepackage{subfigure}
\usepackage{psfrag}

\begin{document}

\title{Surface-hopping dynamics and decoherence with quantum equilibrium structure}
\author{Robbie Grunwald}
\email{robbie.grunwald@chem.utoronto.ca}
\affiliation{Chemical Physics Theory Group, Department of
Chemistry, University of Toronto, Toronto, ON M5S 3H6 Canada}

\author{Hyojoon Kim}
\email{hkim@chem.utoronto.ca}
\affiliation{Department of Chemistry, Dong-A University, Hadan-2-dong, Busan 604-714 Korea}

\author{Raymond Kapral}
\email{rkapral@chem.utoronto.ca}
\affiliation{Chemical Physics Theory Group, Department of
Chemistry, University of Toronto, Toronto, ON M5S 3H6 Canada}

\date{\today}

\begin{abstract}
In open quantum systems decoherence occurs through interaction of a quantum subsystem with its environment. The computation of expectation values requires a knowledge of the quantum dynamics of operators and sampling from initial states of the density matrix describing the subsystem and bath.  We consider situations where the quantum evolution can be approximated by quantum-classical Liouville dynamics and examine the circumstances under which the evolution can be reduced to surface-hopping dynamics, where the evolution consists of trajectory segments evolving exclusively on single adiabatic surfaces, with probabilistic hops between these surfaces. The justification for the reduction depends on the validity of a Markovian approximation on a bath averaged memory kernel that accounts for quantum coherence in the system. We show that such a reduction is often possible when initial sampling is from either the quantum or classical bath initial distributions.  If the average is taken only over the quantum dispersion that broadens the classical distribution, then such a reduction is not always possible.  
\end{abstract}

\maketitle

\section{Introduction}\label{intro}

The exact quantum dynamics of processes such as proton transfer in chemical or biological systems,~\cite{bell73, kornyshev97, hadzi98} vibrational energy relaxation in condensed phases,~\cite{everitt:4467, skinner99} or ultrafast dynamics of photoexcited molecular systems~\cite{michl90, mukamel05} cannot be computed directly, due to the exponential scaling of computational time with the number of degrees of freedom of the system.  Consequently, a number of different approximate quantum dynamical schemes have been used to circumvent this difficulty.  Surface-hopping schemes,~\cite{tully71, tully90, hammesschiffer94} where classical degrees of freedom evolve on single adiabatic surfaces and make probabilistic hops from one surface to another, are the most commonly used such methods.  Generally, one would expect the system to evolve into a coherent state as a result of the coupling between adiabatic states.  The justification for the restriction of the evolution to single adiabatic surfaces is based on the fact that interaction with the environment rapidly destroys coherence resulting in the collapse of the system onto a single adiabatic surface.  Several prescriptions have been put forth to capture this physics.~\cite{tully90, hammesschiffer94, rossky&webster&wang&friesner94, rossky&prezhdo97, rossky&bittner95, coker95, mei96}

Mixed quantum-classical Liouville dynamics~\cite{alek81, geras82, mukamel94, martens97, kapral99, santer01, horenko02, shi04} can be used to model quantum dynamics coupled to a classically evolving environment, and naturally accounts for nonadiabatic transitions and quantum coherence.  Recently we showed how this theory could be recast into the form of a master equation, so that the dynamics involves classical trajectory segments on single adiabatic surfaces, interrupted by hops between these surfaces -- dynamics akin to that in surface-hopping schemes.~\cite{grunwald07} In this formulation, the transition probabilities are determined by coherent evolution trajectory segments where two adiabatic surfaces are coupled.  Decoherence, which is again attributed to interaction with the environment, is accounted for by averaging the coherent evolution segments over an initial distribution of 
the bath degrees of freedom.  

This analysis was applied to the computation of the quantum reactive flux correlation function and rate constant. The reaction rate coefficient was computed in an approximation where the reaction coordinate and bath equilibrium distributions were taken to be classical.  Here we study the effects that result from sampling from quantum equilibrium distributions.  This allows us to study a number of issues related to quantum initial states, decoherence and the validity of simple surface-hopping schemes.

The outline of this article is as follows.  In the next section we present a brief overview of the reduction of the quantum-classical Liouville equation to a master equation description, along with a discussion of the computation of off-diagonal matrix elements of a bath averaged operator in terms of the evolution of the diagonal elements.  In Sec.~\ref{rate-calc} we consider a model for a reactive barrier crossing process in order to illustrate the effects of decoherence on the reaction rate using classical and quantum initial sampling.  We compare the results of master equation and quantum-classical Liouville dynamics.  The quantum initial distribution can be written in terms of a classical distribution broadened by quantum dispersion. In Sec.~\ref{qd} we investigate whether averages over the quantum dispersion are sufficient to justify a Markovian description of the dynamics. Finally, we present our conclusions.

\section{Theory}\label{qme}

The average value of a quantum mechanical operator is given by
\begin{eqnarray}
   \overline{A(t)} & = & {\rm Tr} \hat{A} \hat{\rho}(t) ={\rm Tr} \hat{A}(t) \hat{\rho}(0), \label{avg value}
\end{eqnarray}
where $\hat{A}(t)$ is some observable whose time evolution is given by the Heisenberg equation of motion and $\hat{\rho}(0)$ is the initial quantum density matrix.  Depending on the problem of interest, it may be convenient to represent a subset of the degrees of freedom in some basis. In this study we find it convenient to use an adiabatic basis but any other basis may be chosen. The remaining degrees of freedom are Wigner transformed,~\cite{wigner32} providing a phase-space like representation of this part of the system. The choice of which degrees of freedom are Wigner transformed is based on a consideration of the problem of interest. Thus, an equivalent
exact quantum expression for the average value in Eq.~(\ref{avg value}) is
\begin{eqnarray}
   \overline{A(t)} & = & {\rm Tr'} \int dX \hat{A}_W(X, t) \hat{\rho}_W(X), \nonumber \\
   & = & \sum_{\alpha \alpha'} \int dX A_W^{\alpha \alpha'} (X, t) \rho_W^{\alpha' \alpha} (X),
   \label{adiabatic average}
\end{eqnarray}
where the adiabatic states and energies are given by the solutions to the eigenvalue problem, $\hat{h}_W (R) |\alpha; R \rangle = E_\alpha (R) |\alpha; R \rangle$, with $\hat{H}_W = \hat{h}_W (R) + \frac{P^2}{2 M}$.  In these expressions, $X$ denotes phase space variables $(R, P)$,  ${\rm Tr}'$ denotes the partial trace over the remaining quantum degrees of freedom and the subscript $W$ denotes the partial Wigner transform over the selected degrees of freedom. The evaluation of this expression involves sampling from an initial quantum distribution followed by the evolution of the dynamical variable.

Here, as in previous studies~\cite{kim05, kim07} we assume that the full quantum evolution can be replaced by mixed
quantum-classical Liouville evolution given by~\cite{kapral99, kapral:2006}
\begin{equation}
   \frac{d }{d t}A_W^{\alpha \alpha'}(X, t) = \sum_{\beta \beta'} i \mathcal{L}_{\alpha \alpha', \beta \beta'}
   A_W^{\beta \beta'}(X,   t). \label{Q-C Liouville}
\end{equation}
The quantum-classical Liouville super-operator, $\mathcal{L}_{\alpha \alpha', \beta \beta'}$ is,
\begin{equation}
    i\mathcal{L}_{\alpha \alpha', \beta \beta'} = i (\omega_{\alpha \alpha'} + L_{\alpha \alpha'}) \delta_{\alpha \beta} \delta_{\alpha' \beta'} - \mathcal{J}_{\alpha \alpha', \beta \beta'}\;.\label{Q-C superoperator}
\end{equation}
The two operators in the first term on the right hand side are the frequency corresponding to the energy gap, $\omega_{\alpha \alpha'} = (E_{\alpha} - E_{\alpha'})/\hbar$,  and classical Liouville operator $iL_{\alpha \alpha'} = \frac{P}{M} \cdot \frac{\partial}{\partial R} + \frac{1}{2} (F_W^{\alpha} + F_W^{\alpha'}) \cdot \frac{\partial}{\partial P}$, respectively.  The Hellmann-Feynman force for state $\alpha$ is $F_W^{\alpha}$.  The last term, $\mathcal{J}_{\alpha \alpha', \beta \beta'}$ is the operator responsible for nonadiabatic transitions and corresponding momentum adjustments which ensure energy conservation.  The complete definitions of these operators can be found in Refs.~[\onlinecite{kapral99}] and [\onlinecite{grunwald07}].

We define the subsystem as those degrees of freedom that are of primary interest, while the bath comprises the remainder of the system. For example, in the study of proton transfer~\cite{hanna05}, the subsystem was chosen to be the protonic degree of freedom, $\hat{q}$, plus the solvent polarization, while the bath consisted of the remaining solvent and molecular complex degrees of freedom.  Letting  $X_0 \equiv (R_0, P_0)$ and $X_b \equiv (R_b, P_b)$ be the variables in the subsystem and bath, respectively, that are Wigner transformed, the Hamiltonian can be written as, $\hat{H}_W = \hat{H}_0 + H_{b(0)}$, where the subsystem and bath-plus-interaction Hamiltonians are given by,
\begin{eqnarray}
   \hat{H}_0 & = & \frac{\hat{p}^2}{2m} + \frac{P_0^2}{2 M_0} +\hat{V}_s(\hat{q}, R_0), \label{subsystem H} \\
   H_{b(0)} & = & \frac{P_b^2}{2 M} + V_{b0}(R_b; R_0). \label{bath H + coupling}
\end{eqnarray}
For this Hamiltonian, the adiabatic states depend upon the subsystem coordinates, $R_0$, while the adiabatic energies,
$E_\alpha (R)$, depend upon the full set of coordinates.

The subsystem density matrix, $\hat{\rho}_s (X_0)$, and the bath density, $\hat{\rho}_c (X_b; X_0)$, conditional upon the subsystem configuration, $R_0$, are defined as
\begin{eqnarray}
   \hat{\rho}_s (X_0) & \equiv & \int dX_b \hat{\rho}_W(X_0, X_b), \label{subsystem density} \\
   \hat{\rho}_c(X_b; X_0) & \equiv & \hat{\rho}_W (X_0, X_b) \hat{\rho}_s^{-1} (X_0). \label{conditional density}
\end{eqnarray}
We can express the initial quantum density in the adiabatic basis as,
\begin{eqnarray}
   \rho_W^{\alpha \alpha'} (X_0, X_b) & = & \sum_{\beta} \rho_c^{\alpha \beta} (X_b; X_0) \rho_s^{\beta \alpha'} (X_0). \label{adiabatic density}
\end{eqnarray}
Using this definition, we may rewrite Eq.~(\ref{adiabatic average}) as, 
\begin{eqnarray}
   \overline{A (t)}  & = & \sum_{\alpha \alpha'} \int dX_0 \langle \hat{A}_W (X, t) \rangle_b^{\alpha \alpha'}
   \rho_s^{\alpha' \alpha} (X_0). \label{final average}
\end{eqnarray}
This expression is now composed of an average of the dynamical variable over the conditional bath distribution,
\begin{equation}
   \langle \hat{A}_W (X,t) \rangle_b^{\alpha \alpha'} = \sum_\beta \int dX_b A_W^{\alpha \beta} (X, t) \rho_c^{\beta \alpha'}(X_b; X_0), \label{angled brackets}
\end{equation}
followed by an average over the subsystem density.  The computation of this quantity thus consists of sampling from the initial subsystem quantum distribution, and evolution of the bath averaged dynamical variable.  In the Schr\"odinger picture, one would expect that averaging the density matrix over the bath distribution would cause the coherences, the off-diagonal elements of the subsystem density matrix, to decay rapidly.  This idea served as motivation to derive a master equation, which provided a simple surface-hopping description of the dynamics.

\subsection{Quantum-classical master equation}\label{qcme}

The quantum-classical master equation was derived by writing Eq.~(\ref{Q-C Liouville}) as a coupled set of evolution equations for the diagonal and off-diagonal elements.  By formally solving this set of coupled differential equations, one obtains a generalized master equation for the evolution of the diagonal elements of the operator,~\cite{grunwald07}
\begin{eqnarray}
   &&   \frac{d }{d t}A_d^{\alpha} (X, t) = i L_{\alpha} A_d^{\alpha} (X, t) \label{gme} \\
   && \qquad + \int_0^t dt' \Big( \sum_{\beta} M_{\alpha \beta}^{\alpha \beta}(X,t') A_d^\beta(\bar{X}^{\alpha \beta}_{0 \alpha \beta,t'}, X_{b,t'}, t-t') \nonumber \\
   & & \quad \qquad + \sum_{\nu}M_{\alpha \nu}^{\nu \alpha}(X,t') A_d^\alpha (\bar{X}^{\nu \alpha}_{0\alpha \nu , t'}, X_{b,t'}, t-t') \Big) \;, \nonumber
\end{eqnarray}
where the memory function, $M_{\alpha \beta}^{\alpha \beta}(X,t')$ is defined as,
\begin{equation} \label{mk_function}
   M_{\alpha \beta}^{\alpha \beta}(X,t) = 2{\rm Re} \left[ \mathcal{W}_{\alpha \beta}(t',0)\right] D_{\alpha \beta}(X_0) D_{\alpha \beta}(\bar{X}_{0 \alpha \beta, t'}).
\end{equation}
Here $D_{\alpha \beta}(X_0) = (P_0/M_0) \cdot d_{\alpha \beta}(R_0)$, the nonadiabatic coupling matrix is $d_{\alpha \beta}(R_0) = \langle \alpha; R_0 | \nabla_{R_0} | \beta; R_0 \rangle$, $\mathcal{W}_{\alpha \beta}$ is a phase factor, and the subscripts and superscripts on the memory function label the indices on the first and second $D$ functions respectively.  For the remainder of this article, superscripts and subscripts that appear on variables denote evolution along a particular adiabatic surface, $(\alpha \alpha')$, and the appearance of both a subscript and superscript on a variable indicates the action of two consecutive nonadiabatic transitions.  A bar appearing over $X_0$, denotes a shift in the momentum coming from the $\mathcal{J}$ operator.  The shift is applied to the momentum
coordinate, $P_0$, such that,~\cite{kapral:2006}
\begin{eqnarray}
   \bar{P}_0 & = & P_0+\hat{d}_{\alpha \beta} \Big({\rm sgn}(P_0 \cdot \hat{d}_{\alpha \beta}) \label{eq:deltaP} \\
   & &  \times \sqrt{( P_0 \cdot \hat{d}_{\alpha \beta})^2 + \Delta E_{\alpha \beta}M_0} - (P_0 \cdot \hat{d}_{\alpha \beta})\Big) ,  \nonumber
\end{eqnarray}
and $\bar{X}_0 \equiv (R_0, \bar{P_0})$.  Given that the operator is initially diagonal, the evolution equation (\ref{gme}) is formally equivalent to the mixed quantum-classical Liouville equation,~\cite{kapral99} only now it is organized in two parts: evolution on single adiabatic surfaces and the memory kernel containing all of the off-diagonal evolution.

A generalized master equation was derived by projecting Eq.~(\ref{gme}) onto the subsystem space and making a Markovian approximation on the bath averaged memory kernel that accounts for the coherence in the system. The bath average provides a mechanism for decoherence and leads to decay of the memory kernel. The resulting equation is still non-Markovian in character as a result of the projection of the adiabatic evolution onto the subsystem. The final master equation is obtained by lifting the equation  back into the full phase space to recover a
fully Markovian master equation description:
\begin{eqnarray}
   & & \frac{d}{d t} A_d^\alpha (X,t) = i L_\alpha A_d^\alpha(X,t) \label{Markovian ME} \\
   & &  + \sum_\beta  m_{\alpha \beta}(X_0)  j_{\alpha \to \beta}A_d^\beta (X, t) - m_{\alpha \alpha}(X_0) A_d^\alpha (X, t)  \;. \nonumber
\end{eqnarray}
where
\begin{eqnarray}
   m_{\alpha \beta}(X_0) & = & \int_0^\infty dt' \langle M_{\alpha \beta}^{\alpha \beta} (X, t') \rangle_b \;, \label{m-ab}
\end{eqnarray}
and $j_{\alpha \to \beta}$ is a momentum shift operator.~\cite{grunwald07}  Since Eqs.~(\ref{adiabatic average}) and (\ref{final average}) are equivalent, computation of the average value using either Eq.~(\ref{adiabatic average}) and the master equation after the lift to the full phase space or Eq.~(\ref{final average}) with the non-Markovian subsystem equation will yield the same result.  Note that the subscripts on the second memory term in Eq.~(\ref{Markovian ME}) are the same. This term arises from the memory function corresponding to $\langle M_{\alpha \nu}^{\nu \alpha}(X, t) \rangle_b$. Trajectories accounted for by this term jump to the mean surface and then return to their original surface.  Thus, the net effect is no jump, but a phase factor is introduced.

The master equation~(\ref{Markovian ME}) provides a trajectory description that prescribes evolution on single adiabatic surfaces interspersed with quantum transitions between them.  The probability of a transition is obtained entirely from the coherent evolution.  Furthermore, decoherence is accounted for by the decay of the off-diagonal evolution segments resulting from averaging over the bath distribution.

\subsection{Off-diagonal elements}\label{od_meq}

The expression for the average value in Eq.~(\ref{final average}) includes a sum over all of the matrix elements of the observable.  Here we show that the evolution of the off-diagonal matrix elements can be expressed in terms the diagonal matrix elements so that the master equation can be used to approximately compute these contributions.

The off-diagonal part of the quantum-classical Liouville evolution equation~(\ref{Q-C Liouville}) is given by
\begin{eqnarray}
   \frac{d}{d t} A_o(X, t) & = & i \mathcal{L}^o A_o (X, t) + i \mathcal{L}^{o, d} A_d(X, t), \label{off diagonal evolution}
\end{eqnarray}
where the subscripts $d$ and $o$ indicate diagonal and off-diagonal matrix elements respectively.  Formally solving this equation gives
\begin{eqnarray}
   A_o (X, t) & = & e^{i \mathcal{L}^o t} A_o (X, 0) \nonumber \\
   & &  + \int_0^t dt' e^{i \mathcal{L}^o (t')}i \mathcal{L}^{o,d} A_d (X, t-t'). \label{off diagonal A}
\end{eqnarray}
If we assume that the observable is initially diagonal, then the first term vanishes.  Applying the definition of $\mathcal{L}$ given in Eq.~(\ref{Q-C superoperator}), the second Liouville operator in the above expression reduces to $i\mathcal{L}^{o,d} = - \mathcal{J}^{o,d}$.  Substituting into Eq.~(\ref{off diagonal A}) we find
\begin{equation}
   A^{\alpha \alpha'} (X, t) = \int_0^t dt' \sum_{\mu \mu', \beta} \mathcal{U}^o_{\alpha \alpha', \mu \mu'}(X, t') \mathcal{J}_{\mu \mu', \beta} A_d^\beta (X, t-t'), \label{adiabatic off diagonal evolution}
\end{equation}
where $\mathcal{U}^o_{\alpha \alpha', \beta \beta'}(X, t) =\left(e^{i \mathcal{L}^o(X) t} \right)_{\alpha \alpha', \beta \beta'}$ is the off-diagonal propagator responsible for evolution in off-diagonal space.  For a two level quantum system, this propagator is given exactly by
\begin{equation}
  \mathcal{U}^o_{\alpha \alpha', \beta \beta'} (t) = \mathcal{W}_{\alpha \alpha'}(t, 0) e^{i L_{\alpha \alpha'} (X) t} \delta_{\alpha \beta} \delta_{\alpha' \beta'} \;, \label{2ls-prop}
\end{equation}
for $\alpha, \alpha', \beta, \beta' = 1, 2$ and $\alpha \neq \alpha', \beta \neq \beta'$.  This can be verified by using the Dyson formula and noting that, for a two level system, the only off-diagonal propagator matrix elements that contribute to the dynamics are $\mathcal{U}^o_{12, 12} = \mathcal{U}^{o*}_{21, 21}$.  This definition is generally applicable for weak nonadiabatic coupling.  Acting on the observable with this propagator and using the definition~\cite{kapral99, grunwald07} of $\mathcal{J}$ the evolution equation becomes
\begin{eqnarray}
   A^{\alpha \alpha'}(X, t) =  \int_0^t dt' M_o^{\alpha \alpha'}(X_0, t') \mathcal{A}^{\alpha \alpha'}(X, t, t'), \label{exact evolution-od}
\end{eqnarray}
where we have introduced the off-diagonal memory function,
\begin{eqnarray}
   M_o^{\alpha \alpha'}(X_0, t) & = & \mathcal{W}_{\alpha \alpha'}(t) D_{\alpha \alpha'}(X_{0, t'}^{\alpha \alpha'}), \label{off-diagonal memory kernel}
\end{eqnarray}
and
\begin{eqnarray}
   \mathcal{A}^{\alpha \alpha'} (X, t, t') & = & \left(A_d^{\alpha'} (\bar{X}_{0, t'}^{\alpha \alpha'}, X_{b, t'}, t - t') \right. \\
   & & \left. \hspace{0.5cm} - A_d^{\alpha} (\bar{X}_{0, t'}^{\alpha \alpha'}, X_{b, t'}, t - t') \right). \nonumber \label{notation}
\end{eqnarray}
The expression given in Eq.~(\ref{exact evolution-od}) prescribes evolution along the mean surface for a time $t'$ followed by a jump to a diagonal surface.  Thus, once again, this memory function is made up entirely of coherent evolution segments which carry phase factors.

The off-diagonal memory function differs from that in Eq.~(\ref{mk_function}) by a single momentum-jump operator.  In the diagonal form, evolution starts on an adiabatic surface, jumps to the mean surface and then jumps to another adiabatic surface; thus, two jump operators are involved.  In contrast, off-diagonal evolution starts on a mean surface and, therefore, requires only one jump to get to an adiabatic surface.

By taking the bath average of Eq.~(\ref{exact evolution-od}) and factoring the average on the right hand side, we obtain an evolution equation for the bath averaged observable in the reduced subsystem phase space:
\begin{eqnarray}
   \langle A^{\alpha \alpha'}(X, t)\rangle_b & = &\int_0^t dt' \langle M^{\alpha \alpha'}_o (X_0, t') \rangle_b
   \langle \mathcal{A}^{\alpha \alpha'}\rangle_b (t, t'). \nonumber \\ 
\label{off-diag_intermediate}
\end{eqnarray}
The arguments which are used to justify this approximation have been given in the Appendix of Ref.~[\onlinecite{grunwald07}].  The bath average of the oscillating phase factor in the memory function once again leads to decay.  If this decay is rapid, it provides justification for making a Markovian approximation,
\begin{eqnarray}
   \langle M^{\alpha \alpha'}_o(X_0,t') \rangle_b &\approx& 2 \left(\int_0^\infty dt' \langle M^{\alpha \alpha'}_o(X_0,t') \rangle_b\right) \delta(t') \nonumber \\
   &\equiv& 2 m^{\alpha \alpha'}_o (X_0) \delta(t')\;,\label{eq:mark}
\end{eqnarray}
so that the expression for the evolution of the off-diagonal matrix elements of an observable is given by
\begin{eqnarray}
   \langle A^{\alpha \alpha'} (X, t) \rangle_b = m^{\alpha \alpha'}_o (X_0) \langle \mathcal{A}^{\alpha \alpha'}(X, t) \rangle_b. \label{od-master equation}
\end{eqnarray}
Lifting the equation back to the full phase space we have 
\begin{eqnarray}
   \lefteqn{A^{\alpha \alpha'} (X, t) = } \label{full phase space off diagonal evolution} \\
   & & m_o^{\alpha \alpha'} (X_0) \left(j_{\alpha \alpha'}(X_0)A_d^{\alpha'}(X, t) - j_{\alpha'  \alpha}A_d^{\alpha} (X, t) \right). \nonumber
\end{eqnarray}
The dynamics of the diagonal terms is given by Eq.~(\ref{Markovian ME}), and the momentum shift operators, $j_{\alpha \alpha'}$, defined in Ref.~[\onlinecite{grunwald07}], have been pulled out of the observable.  We have obtained an expression for the evolution of the off-diagonal matrix elements of an observable entirely in terms of diagonal evolution.  This expression prescribes a transition from the mean to each adiabatic surface at time $t = 0$ with probability $m_o^{\alpha \alpha'}(X_0)$, followed by diagonal
evolution.  The resulting trajectories are the same as those given by the diagonal master equation only the momentum is shifted at time $t = 0$.

Since upward jumps from the mean surface require energy input to the subsystem, one must check that there is sufficient energy in the bath to provide the subsystem with the required energy.  Transitions that satisfy $\Delta E_{\alpha \beta}M_0/(P_0 \cdot \hat{d}_{\alpha \beta})^2 > 1$ are not allowed.  In the case where this requirement is satisfied, the off-diagonal evolution is only a function of the evolution along the lower adiabatic surface.

\section{Reaction Rate and Quantum Sampling}\label{rate-calc}

The master equation theory was used to compute the chemical reaction rate for a process $A \rightleftharpoons B$ assuming that the bath and subsystem reaction coordinate were treated classically.~\cite{grunwald07}  Here, we consider the quantum mechanical expression for the rate coefficient that involves averaging over the quantum equilibrium structure.  This allows us to investigate the implications of quantum equilibrium sampling on decoherence and the quantum-classical master equation approximation to the reaction rate.

The  forward rate constant is given by~\cite{kim052}
\begin{eqnarray}
     k_{AB} (t) & = &\frac{1}{n_A^{eq}} \sum_\alpha \sum_{\alpha' \geqslant \alpha}(2 - \delta_{\alpha' \alpha}) \label{rateconst} \\
   & & \int dX \mbox{Re} \left[ N_B^{\alpha \alpha'}  (X, t) W_A^{\alpha' \alpha} (X, \frac{i \hbar \beta}{2}) \right], \nonumber
\end{eqnarray}
where $W_A^{\alpha \alpha'} (X, \frac{i \hbar \beta}{2})$ is the spectral density function that depends on the quantum equilibrium structure, and $N_B^{\alpha \alpha'} (X, t)$ is the time evolved matrix element of the number operator for the product state B. To calculate the rate, one samples initial configurations from quantum equilibrium distributions, and then computes the evolution of the number operator for product state B.

The reaction model we consider here is the same as used in our previous study.~\cite{grunwald07}  The subsystem consists of a two-level quantum system bilinearly coupled to a quartic oscillator and the bath consists of $\nu - 1 = 300$ harmonic oscillators bilinearly coupled to the non-linear oscillator but not directly to the two-level quantum system.  The Hamiltonian and system parameters are the same as those used in prior studies,~\cite{grunwald07, kim06} except the potential parameters are $a=0.25$, and $b=1.0$, $\beta = 1/k_B T$ is taken to be $\beta = 1.0$ and $\beta = 0.5$ corresponding to low and high temperatures, respectively, and the nonadiabatic coupling strength is $\gamma_0 = 1.25$.  The reaction coordinate is $R_0$ and the product species operator $N_B^{\alpha \alpha'} (R_0)=\theta(R_0)\delta_{\alpha \alpha'}$ is initially diagonal in the adiabatic basis.  Here $\theta(R_0)$ is the Heaviside function.

For this reaction model, the barrier region of the potential is locally harmonic along the reaction coordinate $R_0$. Consequently, one can construct an approximate expression for the spectral density function that has the factorized form, $\hat{W}_A (X, i \hbar \beta/2) \approx \hat{\rho}_A(X_0) \rho_b^c(X_b; R_0)$, where $\hat{\rho}_A (X_0)$ is the subsystem spectral density function and $\rho_b^c(X_b; R_0)$ is the Wigner transform of the canonical equilibrium density matrix for the bath in the field of the $R_0$ coordinate.~\cite{kim052} Inserting this form of the spectral density function into the rate coefficient expression, the second line of Eq.~(\ref{rateconst}) may be replaced by
\begin{eqnarray}
   \int dX \mbox{Re} \left[ N_B^{\alpha \alpha'}  (X, t) W_A^{\alpha' \alpha} (X, \frac{i\hbar \beta}{2}) \right] = \hspace{1.5cm} \nonumber \\
   \int dX_0 \mbox{Re} \left[ \langle N_B^{\alpha \alpha'}\rangle_b  (X_0, t) \rho_A^{\alpha' \alpha} (X_0, \frac{i \hbar \beta}{2}) \right], \label{second line rate}
\end{eqnarray}
where the angle brackets indicate an average over the conditional equilibrium distribution, $\rho_b^c(X_b; R_0)$.  The computation of the rate constant now involves the bath averaged observable discussed in Sec.~\ref{qme}, allowing us to make use of the formalism discussed above.

\subsection{Simulation Results}\label{sims}

The evolution of the observable $N_B^{\alpha \alpha'}(X, t)$ was computed using both quantum-classical Liouville dynamics and master equation dynamics.  The simulation of quantum-classical Liouville dynamics was carried out using the sequential short-time propagation algorithm~\cite{kapral&mackernan&ciccotti02} in conjunction with the momentum-jump approximation~\cite{kapral:2006, hanna05} and a bound on the observable.~\cite{kapral&mackernan&ciccotti07, hanna05}  The initial positions and momenta of the quartic oscillator were Monte Carlo sampled from the harmonic part of the quantum equilibrium density distribution $\rho_A^{\alpha' \alpha} (X_0, i \hbar \beta/2)$~\cite{kim07} and the bath coordinates were sampled from the quantum equilibrium density for harmonic oscillators with appropriate frequencies.  Simulation of the master equation consisted of two stages: calculation of the memory kernel, followed by the sequential short-time propagation algorithm restricted to single adiabatic surfaces.~\cite{grunwald07}  Quantum transitions were Monte Carlo sampled using the appropriate memory function for the transition probabilities.

\begin{figure}[htbp]
      \includegraphics[width=7cm]{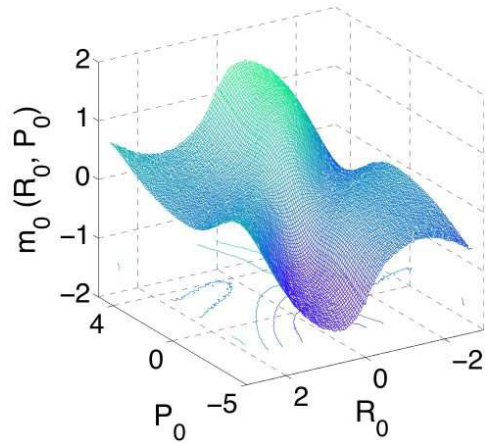}
   \caption{Plot of $m_o (R_0, P_0)$ versus $R_0$ and $P_0$ for $\beta = 1.0$.  }  \label{msurf}
\end{figure}

Computation of the off-diagonal memory function involves a procedure similar to that of the diagonal memory function.~\cite{grunwald07}  A bath configuration is sampled from the quantum equilibrium density for a given subsystem coordinate, $X_0$, and then a trajectory is evolved adiabatically along the mean surface for a prescribed time $t$. The memory function is calculated by numerically integrating Eq.~(\ref{off-diagonal memory kernel}) along the trajectory and then averaging over an ensemble of initial bath configurations. This was done for a range of $X_0$ values resulting in a memory function surface.  The surface generated by integrating this quantity is plotted as a function of $X_0$ in Fig.~\ref{msurf} for the low temperature case.  The surface for the high temperature case is very similar in structure; however, there are minor numerical differences that one would expect to see due to weaker quantum effects at higher temperatures.

Once the memory functions are known, we may calculate the evolution of the observable as discussed elsewhere.~\cite{grunwald07}  The off-diagonal contribution may be calculated using Eq.~(\ref{full phase space off diagonal evolution}).  The initial bath configuration and subsystem coordinate are sampled from the off-diagonal elements of the full quantum equilibrium structure  as before; however, now a nonadiabatic transition from the mean surface to each adiabatic surface is made immediately.  If there is insufficient energy for an upward transition, then there is no population transfer to the excited state for that realization.  Otherwise, the trajectory contributes to the rate, weighted by $m_o^{12} (X_0)$.  The evolution is then computed using the sequential short time propagation algorithm to simulate Eq.~(\ref{Markovian ME}) for each diagonal matrix element.

\begin{figure}[htbp]
   \subfigure{
      \label{rate:1}  %
      \centering
      \psfrag{A}{\large $k_{AB} (t) \times 10^2 $}
      \psfrag{b}{\large a)}
      \includegraphics[width=0.3 \textwidth, angle = -90]{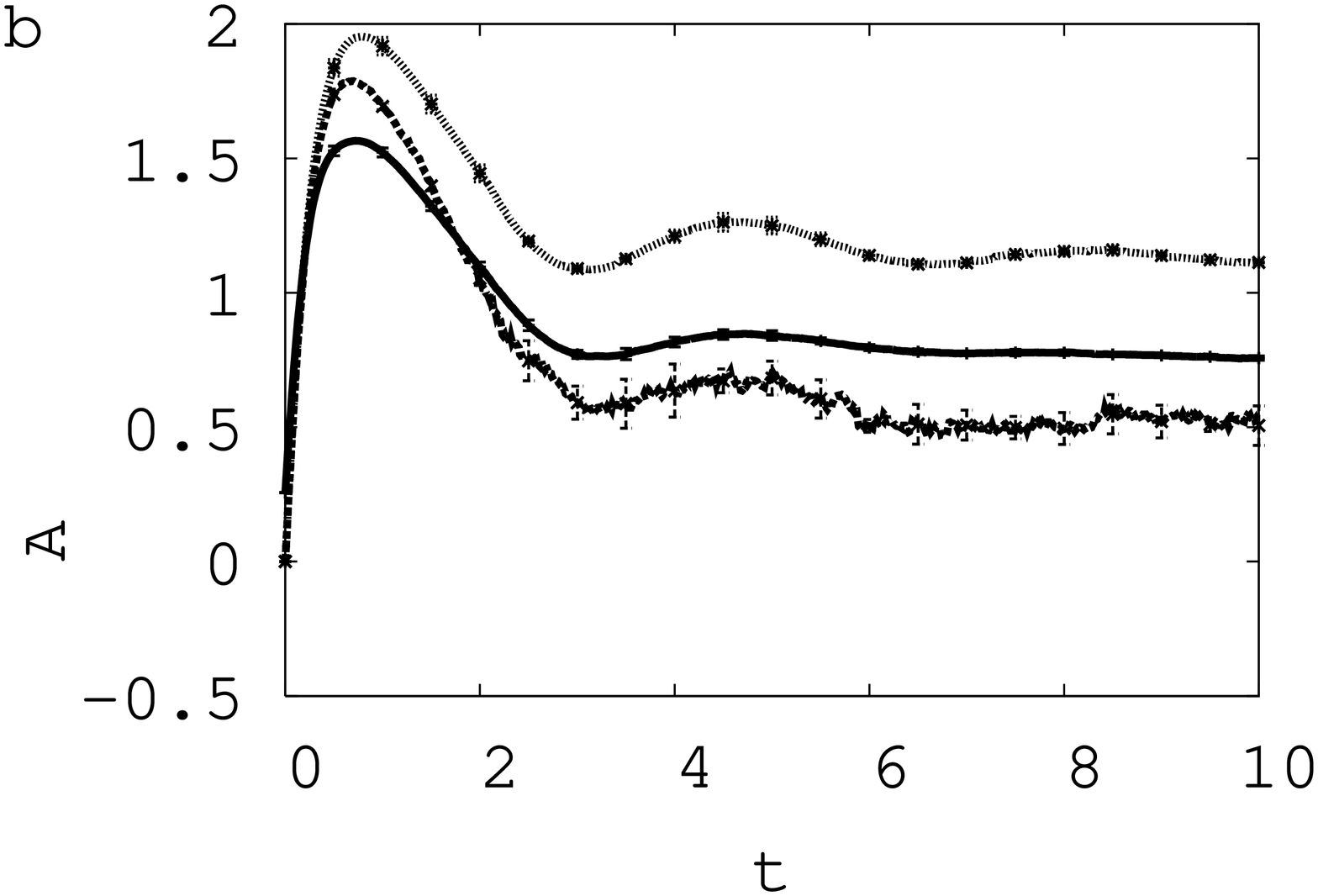}
   }
   \subfigure{
       \label{rate:2}  %
       \centering
       \psfrag{A}{\large $k_{AB} (t) \times 10^2 $}
       \psfrag{b}{\large b)}
       \includegraphics[width=0.3 \textwidth, angle = -90]{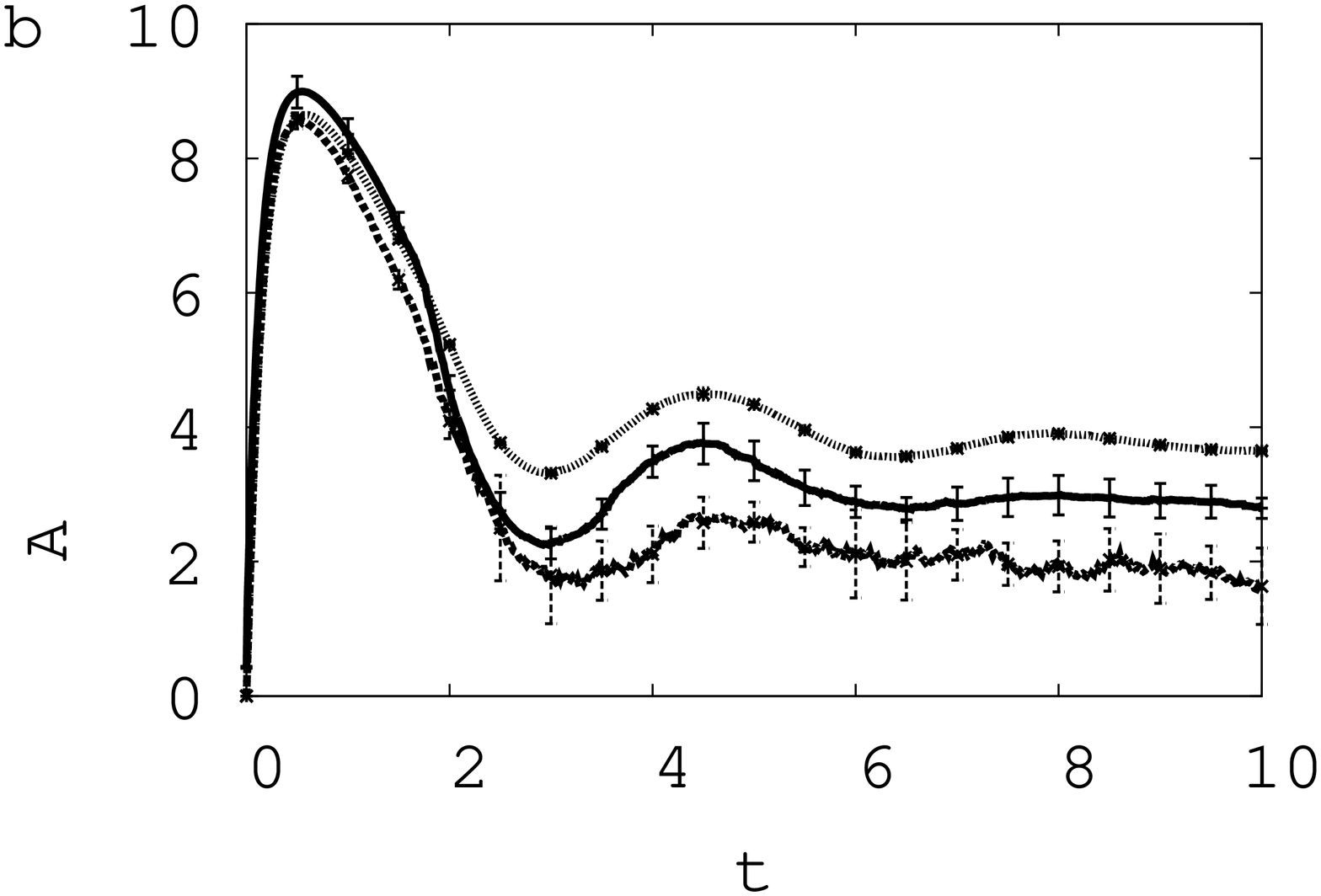}
    }
\caption{Forward rate constant $k_{AB} (t)$ as a function of time.  In both figures, the upper curve is the adiabatic rate, the middle curve is the quantum master equation result and the lowest curve is the mixed quantum-classical Liouville result.  In both panels the microscopic relaxation time $t_{mic}$ is approximately 3.4 units. \subref{rate:1} $\beta = 1.0$.  \subref{rate:2} $\beta = 0.5$.}    
\label{rate} %
\end{figure}

A comparison of the reactive flux correlation function computed using quantum-classical Liouville dynamics, master equation dynamics, and adiabatic dynamics can be seen in Fig.~\ref{rate}.  The lowering of the rate due to nonadiabatic effects is notable in both cases.  The master equation reproduces the short time build-up of the quantum correlation function in both cases in spite of the fact that the short time build-up occurs on the same time scale as the decay of the bath averaged memory kernel.  The zero initial value of the correlation function is a consequence of the quantum mechanical treatment of the reaction coordinate.  As expected, the error bars associated with the master equation calculation are smaller than those from the quantum classical Liouville simulation since the master equation calculation does not involve oscillating phase factors.

\begin{figure}[htbp]

   \subfigure{
      \label{comps:1}  %
      \centering
      \psfrag{A}{\large $k_{AB} (t) \times 10^2 $}
      \psfrag{b}{\large a)}
      \includegraphics[width=0.3 \textwidth, angle = -90]{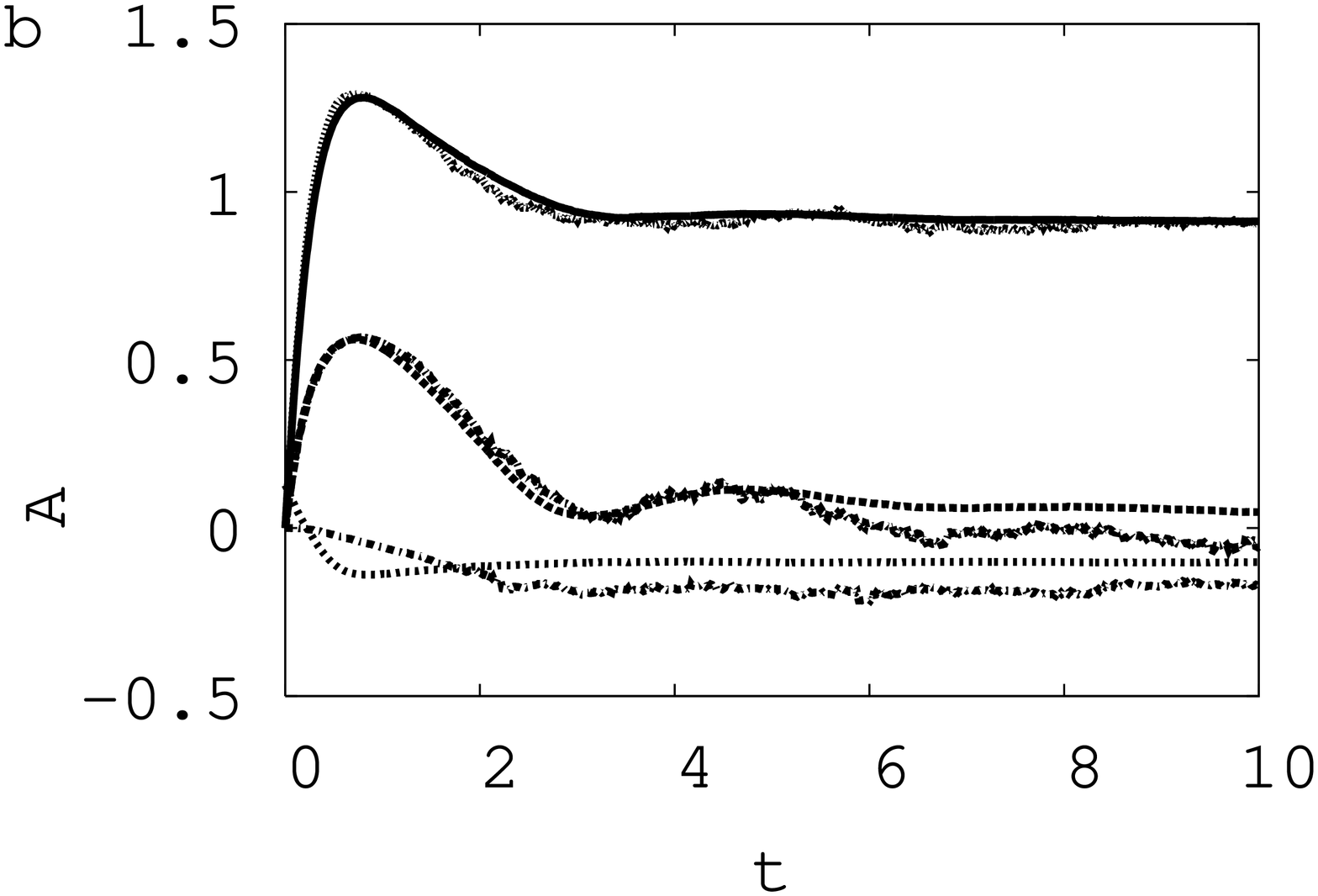}
   }
   \subfigure{
      \label{comps:2}  %
      \centering
      \psfrag{A}{\large $k_{AB} (t) \times 10^2 $}      
      \psfrag{b}{\large b)}
      \includegraphics[width=0.3 \textwidth, angle = -90]{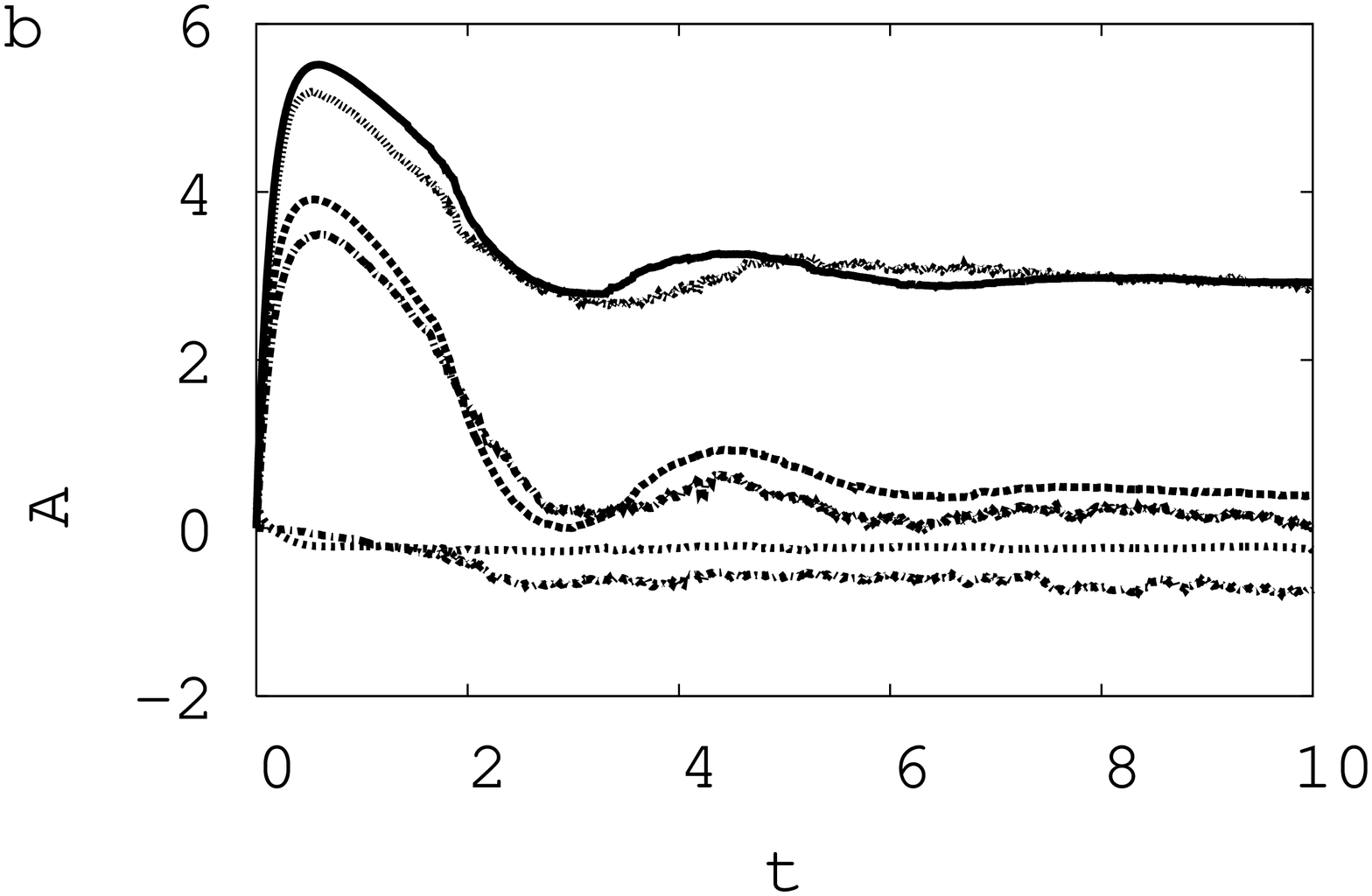}
   }
\caption{Ground state, excited state and off-diagonal contributions to the forward rate constant $k_{AB} (t)$ as a function of time.  The top pair of curves is the ground state contribution, the middle pair is the excited state contribution and the bottom set is the off-diagonal contribution.  In each pair, the smoother curve is the result from the master equation simulation while the noisier curve is from the quantum-classical Liouville calculation.  In both cases it is clear that the discrepancy arises from the disagreement in the off-diagonal and excited state contributions. \subref{comps:1} $\beta = 1.0$.  \subref{comps:2} $\beta = 0.5$.}
 \label{comps}
\end{figure}

The reaction rate constant is given by the plateau values of the correlation function.  The full rate computed with the master equation is close to but larger than the quantum-classical Liouville calculation.  From Fig.~\ref{comps} one sees that the discrepancy is due to the off-diagonal and excited state contributions to the rate.

The quantum-classical Liouville and master equation ground state diagonal contributions to the rate agree very well.  There are discrepancies between the two methods in the excited state and mean surface contributions.  For the excited state contribution, the discrepancy can be understood in terms of recrossing of the barrier region.  In the master equation description, trajectories starting in the excited state, jump down to the ground state, accompanied by an increase of momentum, and rapidly stabilize.  In contrast, for quantum-classical Liouville evolution, trajectories starting in the excited state first make a transition to the mean surface; subsequent evolution results in recrossings on this surface, followed by a transition to the ground state where it stabilizes.  This results in increased recrossing of the barrier region, lowering the plateau value of the correlation function.  Although the ground state contribution is subject to the same qualitative dynamics, upward transitions are associated with a reduction in the momentum and thus stay confined to the barrier region in both descriptions.  As a result, there is a smaller difference between the recrossing corrections for the ground state contribution in the two descriptions.  

\begin{figure}[htbp]
   \includegraphics[width=0.4 \textwidth]{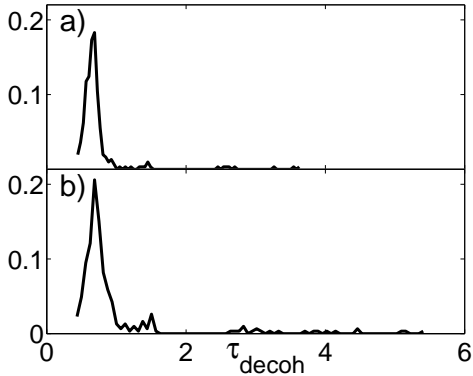}
\caption{Distribution of the decoherence times for the bath averaged memory function at (a) high and (b) low temperatures.}
\label{dtimes}
\end{figure}

The deviations between the off-diagonal terms are largely a consequence of the approximations made in Sec.~\ref{od_meq}.  In particular, the Markov approximation gives rise to a non-zero initial value, which ultimately results in a higher plateau value. Similar effects have been seen in the context of quantum master equations.~\cite{suarez92, gaspard99}  Short time transients typically relax before the slow subsystem processes take place, and the neglect of this behavior leads to a shift in the initial conditions which affects the longer time value of computed quantities.  The relevant time scale to consider in relation to this effect is the decay rate of the memory function or the decoherence time, $\tau_{decoh}$.  The probability distribution of decoherence times was calculated from the first zero crossing of the bath average of the memory function for an ensemble of $X_0$ values.  The results of this calculation can be seen in Fig.~\ref{dtimes}.  The mean decoherence time is slightly longer for the low temperature case with $\tau_{decoh} \approx 0.992$, while the high temperature value is $\tau_{decoh} \approx 0.738$.  The difference can be attributed to the form of the distribution of decoherence times, which exhibits a long tail in the lower temperature case.  
The time scale of the  decay to the plateau value in the off-diagonal contribution to the correlation function is of the same order of magnitude as $\tau_{decoh}$.  Thus, one can infer that the discrepancy appearing in this contribution is due to the neglect of transient memory effects.

The decoherence time as defined above characterizes the decay associated with the bath average of the dynamics of off-diagonal density matrix elements in the adiabatic basis.  Recall that the memory kernel results from substituting the solution of the equation of motion for the off-diagonal density matrix elements into the evolution equation for the diagonal elements.  A number of different measures of the decoherence time have been used in the literature.  A basis-independent quantity that has been used to determine the decoherence time is ${\rm Tr}' \hat{\rho}_s^2$, where $\hat{\rho}_s$ is again the subsystem density matrix averaged over the bath.~\cite{zurek91}  This quantity contains information regarding population transfer in addition to the decay of off-diagonal density matrix elements, resulting in biexponential decay.  The time associated with the fast decay segment of the curve can be identified with the decoherence time.~\cite{foldi01}  We have computed the decoherence time for the high temperature parameter set from the short time decay of ${\rm Tr}'\hat{\rho}_s^2$ and obtained a value of $\tau_{decoh} \approx 0.80$ in agreement with that obtained for the decay of the memory kernel.  We note here that the ratio of decoherence and microscopic time scales $t_{mic}/\tau_{decoh}$, where $t_{mic}$ is the time that characterizes the relaxation  of the correlation function to the plateau value, ranges between 10 and 4 for our simulation results.  In order to provide some insight into these values, we remark that our reaction model provides a simple representation of proton transfer in solution.~\cite{kim06}  In a more realistic model of proton transfer, $t_{mic}$ was found to be approximately 300 fs.~\cite{hanna05}  Using the scaling reported here, our decoherence time would correspond to tens of femtoseconds, a value that is reasonable for condensed phase systems.  A precise assignment of real time units depends on how the model parameters are mapped into those of physical systems.

\section{Quantum Dispersion}\label{qd}

The above results were obtained by sampling from an equilibrium quantum distribution while in our earlier study of decoherence and surface-hopping dynamics~\cite{grunwald07} equilibrium initial states were sampled from a classical initial distribution.  We now explore the differences between the two descriptions.

The quantum mechanical treatment of the reaction coordinate has the most significant effect on the structure of the reactive flux correlation function.~\cite{kim062}  The correlation function rapidly increases from zero, (Fig.~\ref{rate}), and this is qualitatively different from the classical case where the reactive flux correlation function has a non-zero initial value equal to the transition state theory rate.  A quantum mechanical treatment of the reaction coordinate has a more significant effect on the reaction rate than a quantum treatment of the bath.~\cite{kim062}  However, we are interested in decoherence, which arises from a bath average for fixed values of the reaction coordinate.  Therefore, it is important to examine the consequences of treating the bath quantum mechanically or classically.

\begin{figure}[htbp]
   \includegraphics[width=0.5 \textwidth]{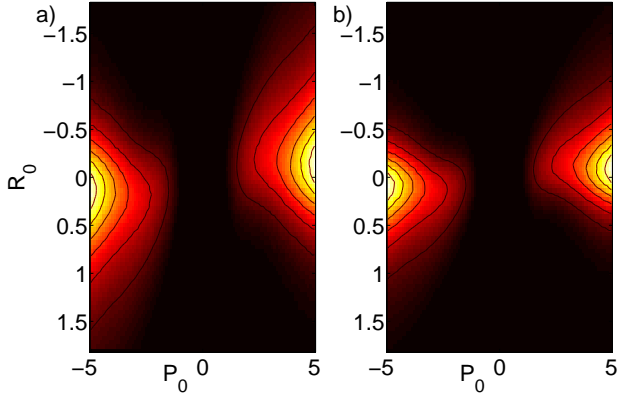}
\caption{Memory surface calculated using (a) the quantum bath equilibrium distribution, and (b) classical bath equilibrium distribution.}
\label{qvsc_mem}
\end{figure}

The memory surface, which is used to compute the nonadiabatic transition rates, is calculated by averaging the memory function over an initial bath distribution.  In Fig.~\ref{qvsc_mem} we compare the surfaces obtained using the quantum and classical bath distributions.
The surface obtained using quantum sampling is somewhat broader than the classical surface.  This is expected since quantum dispersion will broaden the distribution. The magnitude of the variation is quite small and has a small effect on the transition rates.

A different perspective on how the quantum bath structure affects decoherence can be obtained by noting that the quantum equilibrium distribution for a bath of harmonic oscillators can be written as a convolution of the classical density, $\rho_b^{cl} (X_b; R_0) = 1/Z_b^{cl} e^{-\beta H_b}$, with its quantum dispersion function,~\cite{kim062, rossky&hwang04}
\begin{eqnarray}
   \rho_b(X_b; R_0) & = & \int dX'_b g(X'_b - X_b) \rho_b^{cl}(X'_b; R_0),  \label{bath convolution} \\
   g(X_b) & = & \frac{Z^{cl}_b}{Z_b} \prod_j \frac{\beta \omega_j}{2
\pi (u''_j - 1)}\nonumber \\
    && \quad \times   e^{-\frac{\beta}{u''_j - 1} \left( \frac{1}{2M_j} P_j^2 + \frac{1}{2} M_j \omega_j^2 R_j^2 \right)} ,
    \label{quantum dispersion}
\end{eqnarray}
where, $u''_j = u_j \coth u_j$, and $u_j = \beta \hbar \omega_j/2$.  The effect of the quantum dispersion function, $g(X'_b - X_b)$, a Gaussian centered at the bath phase point $X_b$, is to broaden the classical density around each point in phase space.  This function accounts for the quantum fluctuations of the bath.  We consider whether averages over the bath quantum dispersion distribution of the memory kernel for fixed values of the reaction coordinate and bath phase space coordinates yield sufficiently rapid decay of the memory kernel to justify a Markovian approximation.

Starting from the quantum rate expressions~(\ref{rateconst}) and (\ref{second line rate}), we substitute the convoluted form of the bath density to obtain,
\begin{eqnarray}
   k_{AB} (t) & = & \frac{1}{n_A^{eq}} \sum_{\alpha, \alpha'} \int dX \langle N_B^{\alpha \alpha'} (X', t)\rangle_g (X, t) \label{q-disp rate} \\
   & &  \hspace{1.5cm} \times \rho_A^{\alpha' \alpha} (X_0, \frac{i \hbar \beta}{2}) \rho_b^{cl} (X_b; R_0), \nonumber
\end{eqnarray}
where the average value of the species variable over the quantum dispersion is denoted by $\langle N_B^{\alpha \alpha'} (X', t) \rangle_g (X, t) = \int dX'_b g(X'_b - X_b) N_B^{\alpha \alpha'}(X', t)$.  The resulting quantity depends on the subsystem degrees of freedom and bath phase space coordinates determined by the classical bath distribution. An approximate evolution equation for quantum dispersion average of an observable can be obtained a follows.

\begin{figure}[htbp]
   \includegraphics[width=0.3 \textwidth, angle = -90]{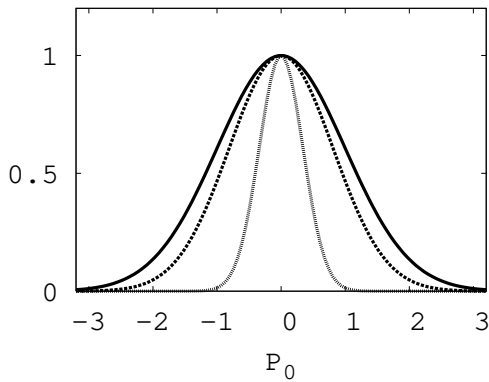}
\caption{Comparison of the classical bath density (solid line) plotted as a function of momentum at $R_0 = 0$ and the quantum dispersion for different oscillator frequencies.  The widest quantum dispersion curve (dashes) corresponds to the cutoff oscillator frequency, $\omega_{max} = 3$, while the narrow curve (dots) corresponds to an intermediate oscillator frequency, $\omega_j = 1.2$.  Low temperature ($\beta = 1$) results are plotted, where one would expect the quantum dispersion to be most significant.  Although the curve corresponding to the maximum oscillator frequency is close to the width of the full bath density, it comprises a small contribution to the quantum dispersion average over the distribution of oscillator frequencies.}
\label{dens}
\end{figure}

Starting from the generalized master equation~(\ref{gme}), we take the average of the entire evolution equation over the quantum dispersion to obtain, 
\begin{eqnarray}
   \lefteqn{\frac{d}{d t} \langle A_d^{\alpha}(X', t) \rangle_g (X, t) = \langle iL_\alpha A_d^{\alpha}(X', t) \rangle_g (X, t) } & &\label{q-disp 1} \\
   & & - \int_0^t dt' \langle M (X', t') A_d (X', t-t') \rangle_g (X, t-t'), \nonumber
\end{eqnarray}
where we have simplified the notation of the memory term analogous to that in Appendix B of Ref.~[\onlinecite{grunwald07}].   Since the factored form of the bath distribution involves a convolution, one cannot use the projection operator formalism that was employed in the derivation of the master equation. Instead, we assume that the observable varies slowly over the width of the quantum dispersion function, so that it may be taken out of the integral. Furthermore, if the observable is approximately constant over the integration region then, $A_d^{\alpha} (X, t) \approx \langle A_d^{\alpha} (X', t) \rangle_g (X, t)$. Using these approximations we obtain
\begin{eqnarray}
   \lefteqn{\frac{d}{d t} \langle A_d^{\alpha}(X', t) \rangle_g (X, t) =} & &   \label{q-disp 2} \\
   & & iL_\alpha \langle A_d^{\alpha}(X', t) \rangle_g (X, t) - \int_0^t dt' \langle M (X', t') \rangle_g (X, t') \nonumber \\
   & & \hspace{3cm} \times \langle A_d (X', t-t') \rangle_g (X, t-t'). \nonumber
\end{eqnarray}
The quantum dispersion average of the memory function appears in the second term of this expression.
\begin{figure}[htbp]
   \subfigure{       
      \label{qdfig:1}  %
      \centering
      \psfrag{A}{$\langle M_{12}^{12} (X', t) \rangle_g (X, t) $}
      \psfrag{b}{\large a)}
      \includegraphics[width=0.3 \textwidth]{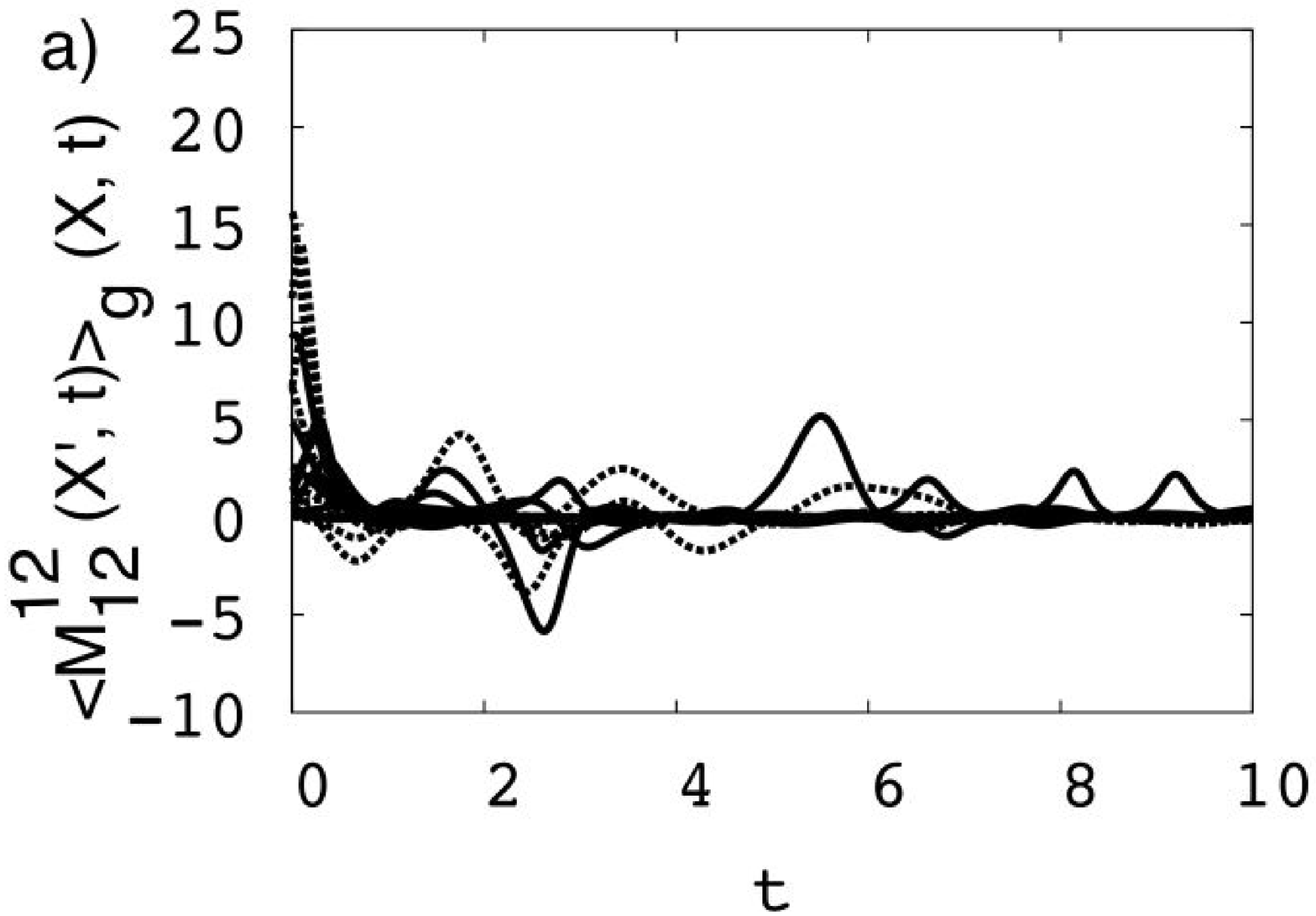}
   }
   \subfigure{
      \label{qdfig:2}  %
      \centering
      \psfrag{A}{ $\langle M_{12}^{12} \rangle_b (X_0, t)$}
      \psfrag{b}{\large b)}
      \includegraphics[width=0.3 \textwidth]{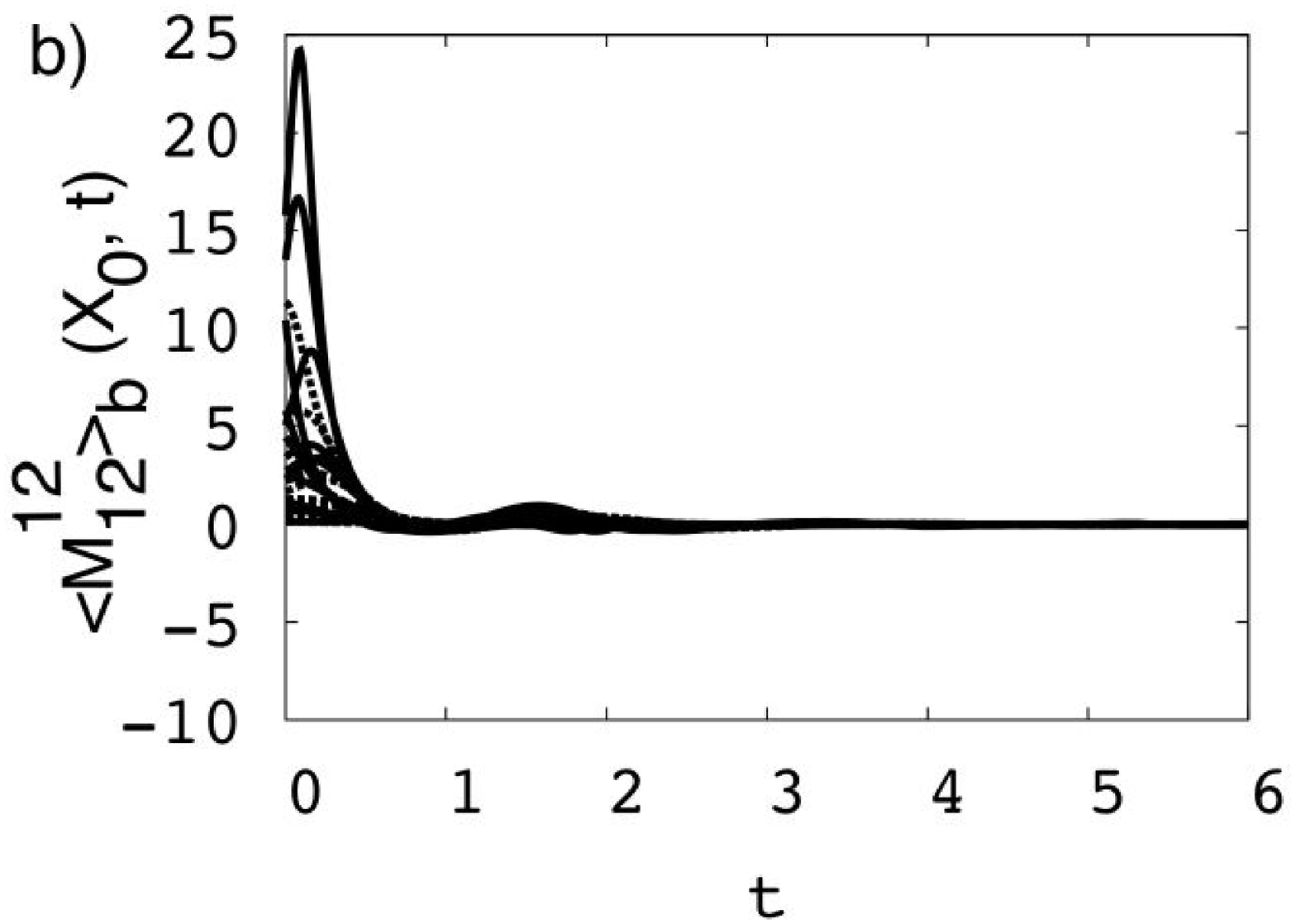}
   }
\caption{Comparison between the quantum dispersion average $\langle M^{12}_{12}(X', t) \rangle_g (X, t)$ and full bath average $\langle M^{12}_{12} \rangle_b (t)$ of the memory function versus time.  \subref{qdfig:1} For high (solid)  and low (dashed) temperatures, the oscillations persist through the simulation time.  \subref{qdfig:2} The full bath average of the diagonal memory function for high (solid) and low (dashed) temperature, for comparison to the quantum dispersion averaging.  Note that the scale on the time axis is shorter and that the function decays quickly.}
\label{qdfig}
\end{figure}
In Fig.~\ref{qdfig:1} we show $\langle M^{\alpha \beta} (X', t) \rangle_g (X, t)$ as a function of time.  Each curve in the plot represents a particular choice of $X_0$ and $X_b$, averaged over the quantum dispersion function $g(X'_b - X_b)$.  From the figure we see that the function does not decay on a short time scale and exhibits long time recurrences.  This structure should be contrasted with that shown in Fig.~\ref{qdfig:2} where a full bath averaged was carried out.  Comparing the classical distribution to the quantum dispersion function (see Fig.~\ref{dens}), one sees that even at low temperatures the range of bath configurations described by $g(X'_b - X_b)$ is narrow.

In this representation, each classical trajectory on the mean surface is replaced by an ensemble of trajectories whose initial values are determined by the quantum dispersion.  The simulation results for our reactive model indicate that each ensemble of trajectories centered at a given system phase space point is confined to a narrow {\em tube} for long times.  Thus, decoherence is not observed on a time scale relevant to the decay of the reactive flux correlation function.  However, if the average is taken over the full bath distribution, then decoherence is rapid, justifying the Markovian approximation leading to master equation dynamics.
\begin{figure}[htbp]
      \subfigure{
             \label{qd_msurf:1}  %
             \centering
                 \includegraphics[width=0.4 \textwidth]{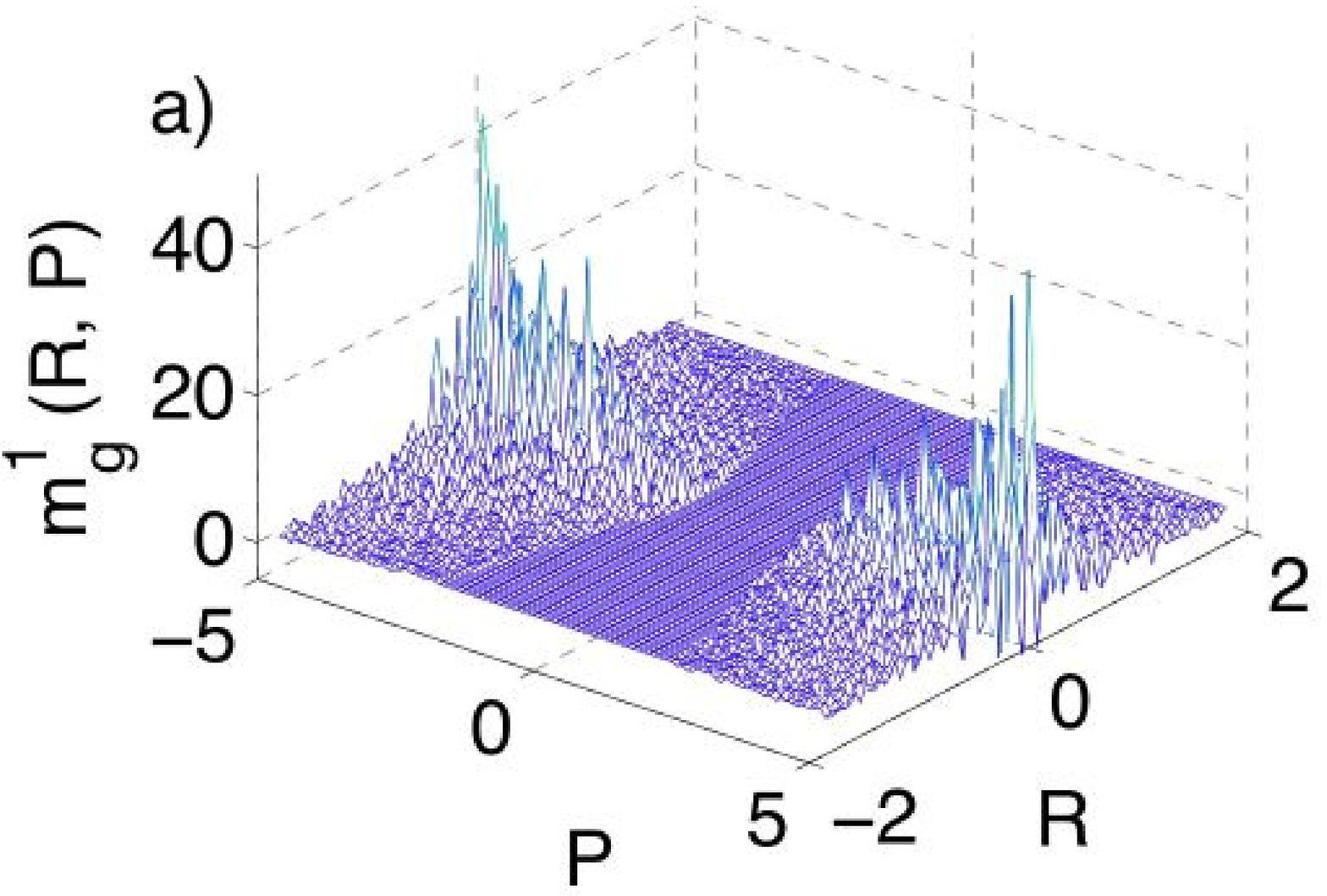}
             }
   \subfigure{
             \label{qd_msurf:2}  %
             \centering
                \includegraphics[width=0.4 \textwidth]{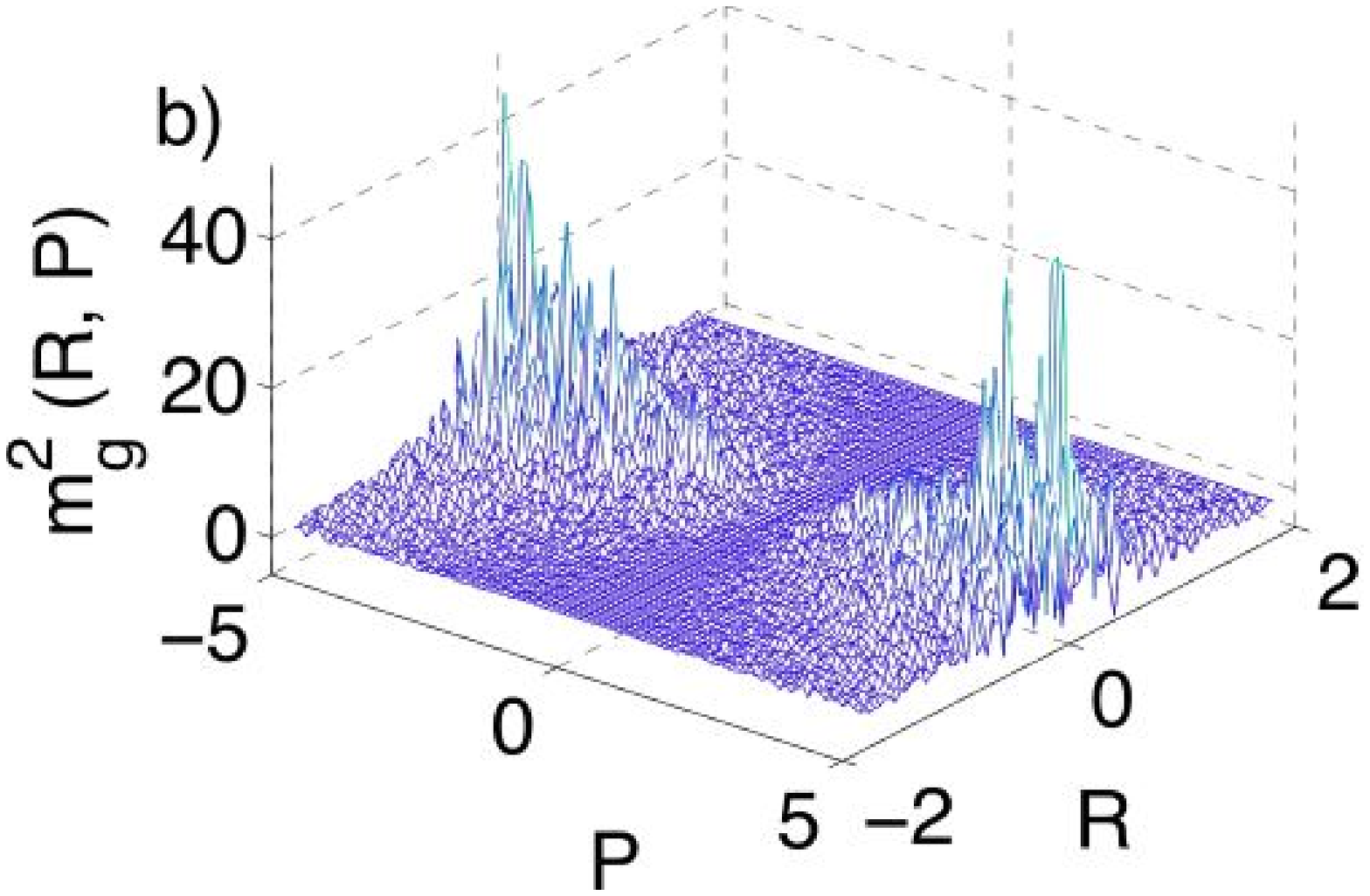}
             }
\caption{The memory surface for $\beta = 1.0$, averaged over the quantum dispersion function using different bath configurations for each subsystem configuration.  The surface is very rough indicating that the choice of initial bath strongly affects the dynamics \subref{qd_msurf:1} Memory surface for upward transitions.  \subref{qd_msurf:2} Memory surface for downward transitions. }
\label{qd_msurf}
\end{figure}



There is another aspect of quantum dispersion versus full bath averaging that merits discussion.  The bath projection led to an equation of motion valid in the subsystem.  The lift to full phase was a device to obtain a fully Markovian description to compute the subsystem expectation values.~\cite{grunwald07} In contrast, if quantum dispersion averaging could be used to justify a Markovian approximation on the memory kernel, one would obtain a master equation valid in the full phase space of the system.  To study the nature of this dependence, we computed the quantum dispersion average of the memory surface as a function of the initial bath configuration.  For each initial subsystem coordinate we sampled a bath configuration from a Boltzmann distribution and then performed an average over the quantum dispersion for that particular configuration.  A new bath configuration was used for each subsystem coordinate.  The results of this calculation, shown in Fig.~\ref{qd_msurf}, indicate that the memory surface is sensitive to the initial bath configuration.  Consequently, it is not possible to reduce the dimensionality of the memory surface in this formulation and this makes the calculation of the transition probabilities difficult.  The surface may be smoothed by averaging over the ensemble of bath configurations, i.e., taking the full bath average.

\section{Conclusion}\label{conc}

Decoherence in open quantum systems has its origin in the interactions of the system with its environment.~\cite{zurek91} This paper focused on the effects of quantum versus classical equilibrium sampling on decoherence and the computation of average values and correlation functions. Both of these quantities can be written in forms that involve quantum initial sampling (or classical sampling when this approximation is appropriate) and quantum dynamics, which we model by quantum-classical Liouville dynamics.  As in our earlier study,~\cite{grunwald07} the decoherence problem can be cast in the form of the validity of a Markovian approximation on the bath averaged memory kernel for the evolution of the diagonal elements of the density or operators.  

The simulations on a simple reaction model with quantum initial sampling allowed us to assess the validity of simple surface-hopping dynamics and how it relates to decoherence.  Since the reaction coordinate is treated quantum mechanically, the reactive flux correlation function starts from zero and finally reaches a plateau value from which the rate constant can be determined.  In spite of the fact that this build-up typically occurs on a similar time scale to the decoherence time, the initial structure of the correlation function is well described by master equation dynamics. 

The role of quantum dispersion on decoherence was also investigated.  Depending on system parameters, averages over the quantum dispersion around fixed bath phase space coordinates may be insufficient to provide a justification for a Markovian approximation to the memory kernel.  Furthermore, simulations on the reaction model indicated that the quantum dispersion average of the memory kernel is a strong function of the bath coordinates and, thus, so are the transition probabilities in the master equation. 

An aim of this investigation was to examine the conditions under which simple surface-hopping descriptions of the dynamics, where the system evolves on single adiabatic surfaces interrupted by quantum transitions between such surfaces, are valid.  An average of the memory kernel over some distribution that results in decay on a time scale faster than the decay of the correlation function of interest is required in order for a Markovian master equation description of the dynamics to be valid.  The decoherence of the subsystem must occur on a time scale that is shorter than that of slow subsystem processes.  When these conditions are satisfied, the resulting dynamics provides a useful tool to compute expectation values since it is more stable with smaller statistical uncertainties.  The results of this work should prove to be useful for understanding the dynamics of many body quantum systems more complex than those considered for our model calculations.

\section*{Acknowledgements}

This work was supported in part by a grant from the Natural Sciences and Engineering Research Council of Canada.

\end{document}